\documentclass[12pt]{article}

\usepackage{graphicx}
\begin{document}

\begin{center}
{\bf Regular model of magnetized black hole with rational nonlinear electrodynamics} \\
\vspace{5mm} S. I. Kruglov
\footnote{E-mail: serguei.krouglov@utoronto.ca}
\underline{}
\vspace{3mm}

\textit{Department of Physics, University of Toronto, \\60 St. Georges St.,
Toronto, ON M5S 1A7, Canada\\
Department of Chemical and Physical Sciences, University of Toronto Mississauga,\\
3359 Mississauga Road North, Mississauga, Ontario L5L 1C6, Canada} \\
\vspace{5mm}
\end{center}
\begin{abstract}
A modified Hayward metric of magnetically charged black hole space-time based on rational nonlinear electrodynamics with the Lagrangian ${\cal L} = -{\cal F}/(1+2\beta{\cal F})$ is considered. We introduce the fundamental length, characterising quantum gravity effects. If the fundamental length vanishes the general relativity coupling to rational nonlinear electrodynamics is recovered. We obtain corrections to the Reissner$- $ Nordstr\"{o}m solution as the radius approaches infinity. The metric possesses a de Sitter core without singularities as $r\rightarrow 0$. The Hawking temperature and the heat capacity are calculated. It was shown that phase transitions occur and black holes are thermodynamically stable at some event horizon radii. We demonstrate that curvature invariants are bounded and the limiting curvature conjecture takes place.
\end{abstract}

\section{Introduction}

One of problems in General Relativity (GR) is singularities. The Schwarzschild metric for neutral black holes (BHs) and Reisner $-$Nordstr\"{o}m metric for charged BHs have curvature singularities at $r=0$.
To avoid singularities GR has to be modified for the large curvatures. There are different approaches to deform GR. To solve the singularity problem one can assume that there is a critical energy $\mu$ and the corresponding length $l=\mu^{-1}$ so that the metric is modified when the space-time curvature is in the order of $l^{-2}$ \cite{Sakharov}, \cite{Markov}, \cite{Markov1} (see also \cite{Frolov}).
The length scale $l$ characterises quantum gravity effects and this approach is beyond classical GR. The modification
of GR considered here is important at the Planck scale $l_{Pl}$. When curvature invariants are much greater than $l^{-2}$ the deformation of GR,
considered here,  should  be taken into account. Otherwise standard results are recovered. For a neutral BH a simple metric with the
length $l$ was proposed by Hayward \cite{Hayward} in the framework of quantum gravity theory. We follow this avenue by modifying the Hayward metric by the replacement of the constant mass $M$ with the mass $M(r)$ depending on the radius. It is worth noting that the complete quantum gravity theory is not developed yet. The fundamental constant $l$ only mimics the quantum gravity effects at small radii of BHs and huge curvatures within a classical model. Therefore, we do not imply the concrete deformation of the Einstein equations and corresponding equations of motion. Some authors, to avoid the black hole singularities, postulated the metric for the regular BHs but without discussing what are sources of gravity \cite{Bardeen}. The corresponding solutions in these models are not exact solutions to Einstein equations.

In this paper we postulate a new metric with the fundamental length $l$, a la Hayward, which for $l=0$ gives
the regular magnetically charged BH solution of Einstein equations with the source of rational nonlinear electrodynamics (RNED) proposed in \cite{Kr3}. The BH thermodynamics and phase transitions are investigated. The authors of \cite{Ali} and \cite{Halilsoy} also studied BH solutions with the modified Hayward metric based on nonlinear electrodynamics (NED) proposed in \cite{Kr8} and \cite{Kr7}, respectively. In \cite{Kr2} the similar study was performed for a model considered in \cite{Bronnikov}. There are different models of NED known in the literature (not a complete list) \cite{BI0}-\cite{Habib}. It is worth noting that the general form of the Lagrangian density for NED in the case of nonsingular black holes was given in \cite{Bronnikov2}.
Here, we explore RNED which has attractive features such as a simplicity and regularity.
The RNED coupled to GR gives the regular magnetic black holes \cite{Kr}. Also, the size of the shadow of M87* black hole calculated within RNED-GR is in agrement with the data of Event Horizon Telescope Collaboration \cite{Kr1}. In addition, the RNED describes the inflation of the universe in accordance with the astrophysical data \cite{Kr4}.

It should be noted that the Hayward metric corresponds to the neutral black hole and regularises the Schwarzschild solution. The modify Hayward metric, which we propose, describes the magnetically charged black hole with nonzero Schwarzschild mass $m_0$ and magnetic mass $m_M$. The magnetically charged black hole with RNED, in the framework of GR, has a solution which is regular only if $m_0=0$ \cite{Kr3}. The modify Hayward metric, for magnetized black hole, that we discuss here, is regular even for $m_0\neq 0$.

The structure of the paper is as follows. In Sect. 2 we study the regular magnetized BH solution within nonlinear rational electrodynamics. A modified Hayward metric of magnetically charged black hole space-time and the fundamental length, characterising quantum gravity effects, are introduced. Thermodynamics and phase transitions in this model are investigated in Sect. 3. The Hawking temperature and the heat capacity are evaluated. We show that phase transitions take place. In Sect. 4, it is demonstrated that curvature invariants are bounded and the limiting curvature conjecture takes place. In Sect. 5 we make a conclusion.

We use here a natural unit system with $\epsilon_0=\mu_0=c=\hbar=k_B=1$. This yields a single scale value - the length $L$.
In this system the Newton constant $G$ has a unit $[G]=L^2$, the mass $m$ possesses the init $[m]=L^{-1}$ and the charge $q$ is unitless.

\section{A regular magnetized BH solution}

Let us consider the Lagrangian of RNED \cite{Kr3} to obtain the magnetically charged BH solution
\begin{equation}
{\cal L} = -\frac{{\cal F}}{1+2\beta{\cal F}},
 \label{1}
\end{equation}
with the parameter $\beta$ of the dimension of (length)$^4$, ${\cal F}=(1/4)F_{\mu\nu}F^{\mu\nu}=(\textbf{B}^2-\textbf{E}^2)/2$,
$F_{\mu\nu}=\partial_\mu A_\nu-\partial_\nu A_\mu$ is the field tensor. The correspondence principle holds because at the weak fields $\beta {\cal F}\ll 1$, we have the Maxwell limit and the Lagrangian (1) becomes
\begin{equation}
{\cal L}\rightarrow-{\cal F}.
 \label{2}
\end{equation}
We will consider the magnetic BHs because, as was proven in [13], the solution for electrically charged BHs in GR coupled with
NED, which has at weak-field limit the Maxwell electrodynamics, is singular. Here the magnetic monopole is due to space-time curvature and is realized in BHs. We use the spherically symmetric line element squared
\begin{equation}
ds^2=-f(r)dt^2+\frac{1}{f(r)}dr^2+r^2(d\vartheta^2+\sin^2\vartheta d\phi^2).
 \label{3}
\end{equation}
To derive the metric function representing the static magnetic regular BH we explore the Hayward metric function \cite{Hayward}
\begin{equation}
f(r)=1-\frac{2GMr^2}{r^3+2GMl^2},
\label{4}
\end{equation}
with $G$ being the Newton constant, and $l$ is the fundamental length. In a Hayward  model of the neutral BH the mass $M$ is constant. The metric function (4) can be found as a solution in GR coupled with NED, containing the length $l$, $G$ and a nonzero magnetic charge \cite{Kr}. In this case the correspondence principle is broken and at the weak-field limit the NED Lagrangian is not converted to the Maxwell Lagrangian. Thus, the source of gravity within GR is questionable. Therefore, we interpret the Hayward metric, within modified gravity theory, as a metric which takes into account quantum corrections characterized by the fundamental constant $l$. This approach is beyond the GR and $l$ mimics the quantum effects when $r\ll l$. A variable mass $M(r)$ will be substituted into Eq. (4) instead of a constant mass $M$.
Thus, we postulate the metric (4) with the variable mass $M(r)$ and consider such metric in the framework of modified GR by quantum corrections with the fundamental length $l$. Similar procedure was used also in the papers \cite{Ali} and \cite{Halilsoy} for other NEDs.
When $l=0$ in Eq. (4) with a constant mass $M$, we have the Schwarzschild metric of an uncharged BH which is a solution to Einstein's equation. We assume that the BH is magnetically charged and the mass function of the BH varies with $r$ \cite{Bronnikov},
\begin{equation}
M(r)=m_0+\int_0^r\rho(r) r^2dr=m_0+\int_0^\infty\rho(r) r^2dr-\int_r^\infty\rho(r) r^2dr,
\label{5}
\end{equation}
where $m_0$ (the constant of integration) is the Schwarzschild mass and $ m_M = \int_0^\infty\rho(r)r^2dr$ is the magnetic mass of the BH, $\rho(r)$ is the magnetic energy density. It should be noted that we have a regular metric with RNED (1) in GR if $m_ 0=0$ \cite{Kr}. But at a nonzero $m_ 0$ the singularity is present.
When the charge $q=0$ the mass function $M=m_0$ becomes constant and we come to the Hayward metric function (4). The magnetic energy density at $\textbf{E}=0$ is \cite{Kr}
\begin{equation}\label{6}
  \rho(r)=-{\cal L} = \frac{{\cal F}}{1+2\beta{\cal F}}
  =\frac{B^2}{2(\beta B^2+1)}=\frac{q^2}{2(r^4+\beta q^2)},
\end{equation}
where ${\cal F}=B^2/2=q^2/(2r^4)$, and $q$ is a point-like magnetic charge. Making use of Eq. (6) the mass function (5) becomes
\[
M(x)=m_0+\frac{q^{3/2}}{8\sqrt{2}\beta^{1/4}}\biggl(\ln\frac{x^2-\sqrt{2}x+1}{x^2+\sqrt{2}x+1}
\]
\begin{equation}
+2\arctan(\sqrt{2}x+1)-2\arctan(1-\sqrt{2}x)\biggr),
\label{7}
\end{equation}
where we introduce the dimensionless variable $x=r/\sqrt[4]{\beta q^2}$. The BH magnetic mass is given by \cite{Kr}
\begin{equation}\label{8}
 m_M = \int_0^\infty\rho(r)r^2dr=\frac{\pi q^{3/2}}{4\sqrt{2}\beta^{1/4}}\approx 0.56\frac{q^{3/2}}{\beta^{1/4}},
\end{equation}
and the total BH mass is $M(\infty)\equiv m=m_0+m_M$. It is worth noting that in classical electrodynamics the magnetic mass, which is the total magnetic energy of a magnetic monopole is infinite, but in our case the magnetic energy of the magnetized BH is finite.
The metric function of a charged BH becomes
\begin{equation}
f(r)=1-\frac{2GM(r)r^2}{r^3+2GM(r)l^2},
\label{9}
\end{equation}
with $M(r)$ given by (7). At $l=0$ we have the metric function of a charged BH within GR \cite{Kr}.
With the help of Eqs. (7)-(9) we obtain the metric function
\begin{equation}\label{10}
f(x)=1-\frac{x^2(Ag(x)+C)}{4\sqrt{2}x^3+CD+Bg(x)},
\end{equation}
where the dimensionless constants $A$, $B$, $C$, $D$ and the function $g(x)$ are given by
\[
A=\frac{Gq}{\sqrt{\beta}},~~~B=\frac{Gl^2}{\beta},~~~C=\frac{8\sqrt{2}Gm_0}{\sqrt{q}\beta^{1/4}},~~~D=\frac{l^2}{q\sqrt{\beta}},
\]
\begin{equation}\label{11}
~~g(x)=\ln\frac{x^2-\sqrt{2}x+1}{x^2+\sqrt{2}x+1}+2\arctan(\sqrt{2}x+1)-2\arctan(1-\sqrt{2}x).
\end{equation}
Do not confuse the parameter $B$ with the magnetic field.
Making use of Eqs. (10) and (11) one finds the asymptotic of the metric function as $r\rightarrow\infty$ and $r\rightarrow 0$
\begin{equation}\label{12}
  f(r)=1-\frac{2Gm}{r}+\frac{Gq^2}{r^2}+\frac{4G^2ml^2}{r^4}\left(m-\frac{q^2}{r}\right) +
{\cal O}(r^{-6})~~~~r\rightarrow\infty,
\end{equation}
\begin{equation}\label{13}
  f(r)=1-\frac{r^2}{l^2}+\frac{5r^{10}}{G\beta q^4l^4}+{\cal O}(r^{11})~~~~r\rightarrow 0.
\end{equation}
According to Eq. (12) the corrections to the RN solution are in the order of ${\cal O}(r^{-4})$. At $l=0$ Eq. (12) becomes the equation found in \cite{Kr}. As $r\rightarrow \infty$ the spacetime becomes flat. Because $\lim_{r\rightarrow 0}f(r)=1$ the space-time of the BH is regular and possesses a smooth de Sitter core.
The plot of the function $f(x)$ is depicted in Fig. 1. \begin{figure}[h]
\includegraphics[height=3.0in,width=3.0in]{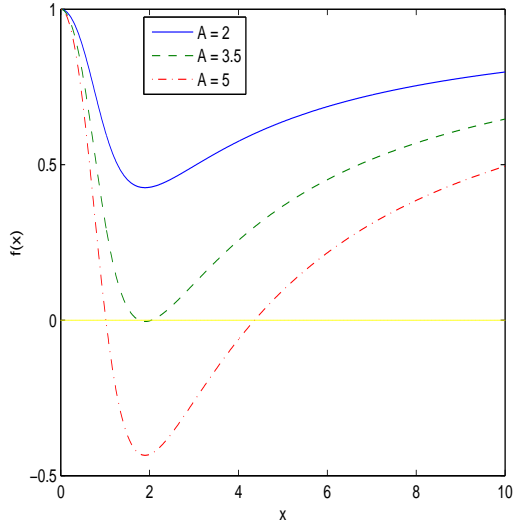}
\caption{\label{fig.1}The plot of the function $f(x)$ for $B=1$ and $C=0$ ($m_0=0$). The dashed-dotted line corresponds to $A=5$, the solid line corresponds to $A=2$ and the dashed line corresponds to $A\approx 3.5$.}
\end{figure}
Figure 1 shows that at $A<3.5$ ($B=1$, $C=0$) horizons are absent and we have the particle-like solution and BHs are absent. At $A\approx 3.5$ we have the extreme solution (one horizon). When $A>3.5$, there are two BH horizons. The plot of the parameter $A$ at $C=0$  versus the horizon radii $x_h$ ($f(x_h)=0$) is in Fig. 2. The inverse function gives the dependence of $x_h$ on the parameter $A$.
\begin{figure}[h]
\includegraphics[height=3.0in,width=3.0in]{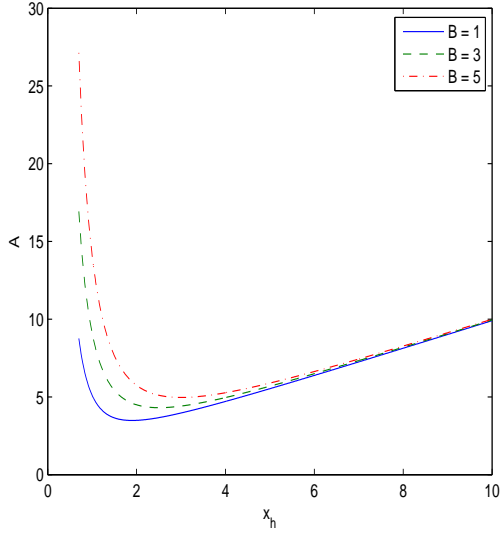}
\caption{\label{fig.2}The plot of the function $A$ versus $x_h$ for $C=0$ ($m_0=0$). The dashed-dotted line corresponds to $B=5$, the solid line corresponds to $B=1$ and the dashed line corresponds to $B=3$.}
\end{figure}

\section{Thermodynamics and phase transitions}

The mass function (7) and lapse function of metric (3) depend only on the radial coordinate, and therefore, the
spacetime is stationary and possesses a timelike Killing vector, $\chi^\mu$, so that $\chi^\mu \partial_\mu=\partial_t$ and $\chi^\mu=(1,0,0,0)$. The covariant Killing vector for the metric (3) is $\chi_\nu=g_{\nu\mu}\chi^\mu=(-f(r),0,0,0)$ so that $\chi_\mu\chi^\mu\rightarrow -1$ as $r\rightarrow\infty$. The surface gravity $\kappa$ for a static BH with a Killing horizon \cite{Wald} is defined as $\chi^\mu\nabla^\nu\chi_\mu=-\kappa\chi^\nu$ (this equation is evaluated at the horizon). Thus, we obtain the surface gravity $\kappa=(1/2)\partial_rf(r)$.
To study the thermal stability of magnetized BHs and the phase transitions we calculate the Hawking temperature  which is given by \cite{Hawking}
\begin{equation}
T_H=\frac{\kappa}{2\pi}=\frac{f'(r)|_{r=r_h}}{4\pi}.
\label{14}
\end{equation}
With the help of Eqs. (10), (14) and the relation $Ag(x_h)x_h^2=4\sqrt{2}x_h^3+Bg(x_h)$ ($f(x_h)=0$) at $C=0$ ($m_0=0$), one finds the Hawking temperature
\begin{equation}
T_H=\frac{1}{4\pi\beta^{1/4}\sqrt{q}}\biggl(-\frac{2}{x_h}
-\frac{4\sqrt{2}x_h^2(4\sqrt{2}x_h^3-3g(x_h)(x_h^4+1))}{g(x_h)(4\sqrt{2}x_h^3+Bg(x_h))(x_h^4+1)}\biggr).
\label{15}
\end{equation}
The dimensionless Hawking temperature $T_H\sqrt{q}\beta^{1/4}$ is depicted in Fig. 3 for $C=0$.
\begin{figure}[h]
\includegraphics[height=3.0in,width=3.0in]{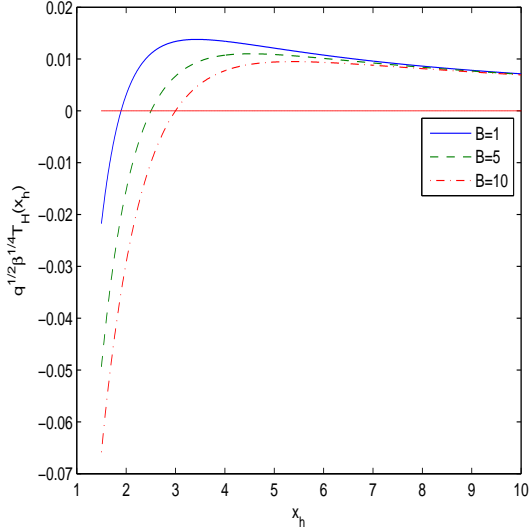}
\caption{\label{fig.3}The plot of the function $\sqrt{q}\beta^{1/4}T_H$ versus horizon radii $x_h$ for $C=0$ ($m_0=0$).  The dashed-dotted line corresponds to $B=10$, the solid line corresponds to $B=1$ and the dashed line corresponds to $B=5$.}
\end{figure}
In accordance with Fig. 3 for the greater value of $B$ (or $l$) the maximum of the Hawking temperature is less. The temperature possesses a maximum corresponding to a phase transition. The similar form of the temperature curve for a BH occurs in the models \cite{Myung1}, \cite{Myung}, \cite{Tharanath}, \cite{Kr5}. One can check that the Hawking temperature (15) with the area law does not satisfy the first law of black
hole thermodynamics. But when $l^{-2}$ is much smaller than curvature invariants the first law of black hole thermodynamics holds. This
occurs if the event horizon radius is not small enough (see Fig. 5). Therefore, we will consider this case in the following to study the
thermodynamics. To describe phase transitions we calculate the heat capacity at the constant charge that is defined by \cite{Novikov}
\begin{equation}
C_q=T_H\left(\frac{\partial S}{\partial T_H}\right)_q=\frac{T_H\partial S/\partial r_h}{\partial T_H/\partial r_h}=\frac{2\pi r_hT_H}{G\partial T_H/\partial r_h},
\label{16}
\end{equation}
where the entropy is $S=A/(4G)=\pi r_h^2/G$. It follows from Eq. (16) that when the Hawking temperature has the extremum, $\partial T_H/\partial r_h=0$, the second-order phase transition takes place because the heat capacity is singular. The plot of the function $GC_q/(\sqrt{\beta}q)$ versus the horizon radius $x_h$ for $C=0$ ($m_0=0$) is depicted in Fig. 4.
\begin{figure}[h]
\includegraphics{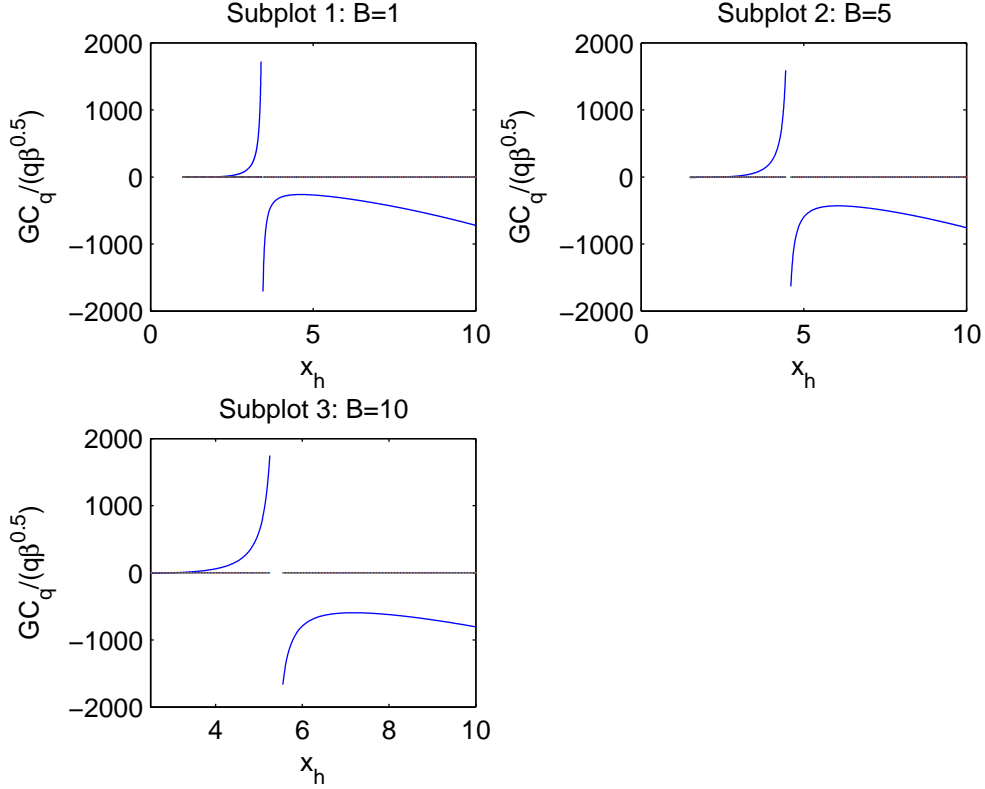}
\caption{\label{fig.4}The plot of the function $C_qG/(\sqrt{\beta}q)$ versus $x_h$ for $C=0$ ($m_0=0$), $B=1$, $B=5$, $B=10$.}
\end{figure}
According to Fig. 4 the second-order phase transitions occur at the discontinuity points where the maximum of the temperature holds. The discontinuity points separate areas with positive and negative heat capacities. The negative capacity correspond to the early stage of the thermodynamics process and the positive heat capacity corresponds to the late stage. When the heat capacity is positive the BH is stable, otherwise the BH is unstable. For the bigger parameter $B$, the second-order BH phase transition occurs at the larger value of the horizon radius $r_h$.
According to Eqs. (15) and (16) at some point, $x_h=x_1$, the Hawking temperature and heat capacity are zero and a first-order phase transition occurs. At this point $x_1$ the BH remnant is formed, where the BH mass is not zero but the Hawking temperature and the heat capacity vanish.  When the maximum of the Hawking temperature $\partial T_H(x_h)/\partial x_h=0$ takes place, at the event horizon radius $x_h=x_2$, the heat capacity possesses the discontinuity and the second-order phase transition holds. Therefore, at the range $x_2>x_h>x_1$ BHs are locally stable. At $x_h>x_2$ the BH evaporates and becomes unstable and at $x_h<x_1$ we have the BH remnant at zero Hawking temperature and heat capacity.

\section{The limiting curvature conjecture}

In a viable theory the curvature invariants are bounded. This statement, named the limiting curvature conjecture, was discussed in \cite{Markov}, \cite{Markov1}, \cite{Frolov}, \cite{Polchinski}. To verify the limiting curvature conjecture we consider the curvature invariants
\[
R(r)=f''(r)+\frac{4f'(r)}{r}-\frac{2(f(r)-1)}{r^2},
\]
\begin{equation}
K(r)\equiv R_{\mu\nu\alpha\beta} K^{\mu\nu\alpha\beta}=
(f''(r))^2+\left(\frac{2f'(r)}{r}\right)^2+\left(\frac{2(f(r)-1)}{r^2}\right)^2,
\label{17}
\end{equation}
where $R(r)$ is the Ricci scalar and $K$ is the Kretschmann scalar. In the terms of the dimensionless variable $x=r/\sqrt[4]{\beta q^2}$ Eq. (17) reads
\[
R(x)q\sqrt{\beta}=f''(x)+\frac{4f'(x)}{x}-\frac{2(f(x)-1)}{x^2},
\]
\begin{equation}
K(x)q^2\beta=(f''(x))^2+\left(\frac{2f'(x)}{x}\right)^2+\left(\frac{2(f(x)-1)}{x^2}\right)^2.
\label{18}
\end{equation}
The Plots of the dimensionless functions $R(x)q\sqrt{\beta}$ and $K(x)q^2\beta$ are depicted in Figs. (5) and (6).
\begin{figure}[h]
\includegraphics[height=3.0in,width=3.0in]{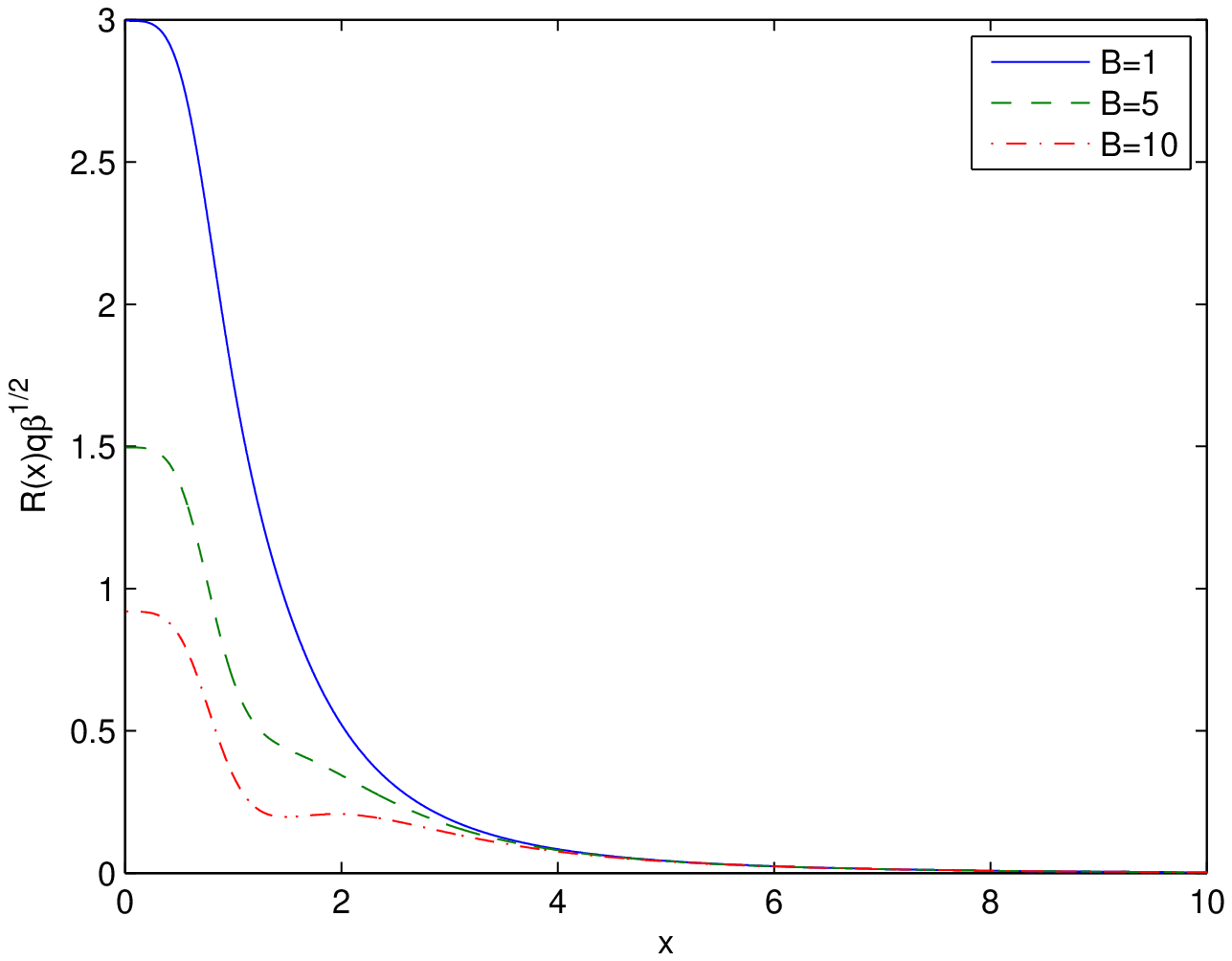}
\caption{\label{fig.5}The plot of the function $R(x)q\sqrt{\beta}$ versus $x$ for $A=1$, $C=0$ ($m_0=0$). The solid line corresponds to $B=1$, the dashed line corresponds to $B=5$ and the dashed-dotted line corresponds to $B=10$.}
\end{figure}
\begin{figure}[h]
\includegraphics[height=3.0in,width=3.0in]{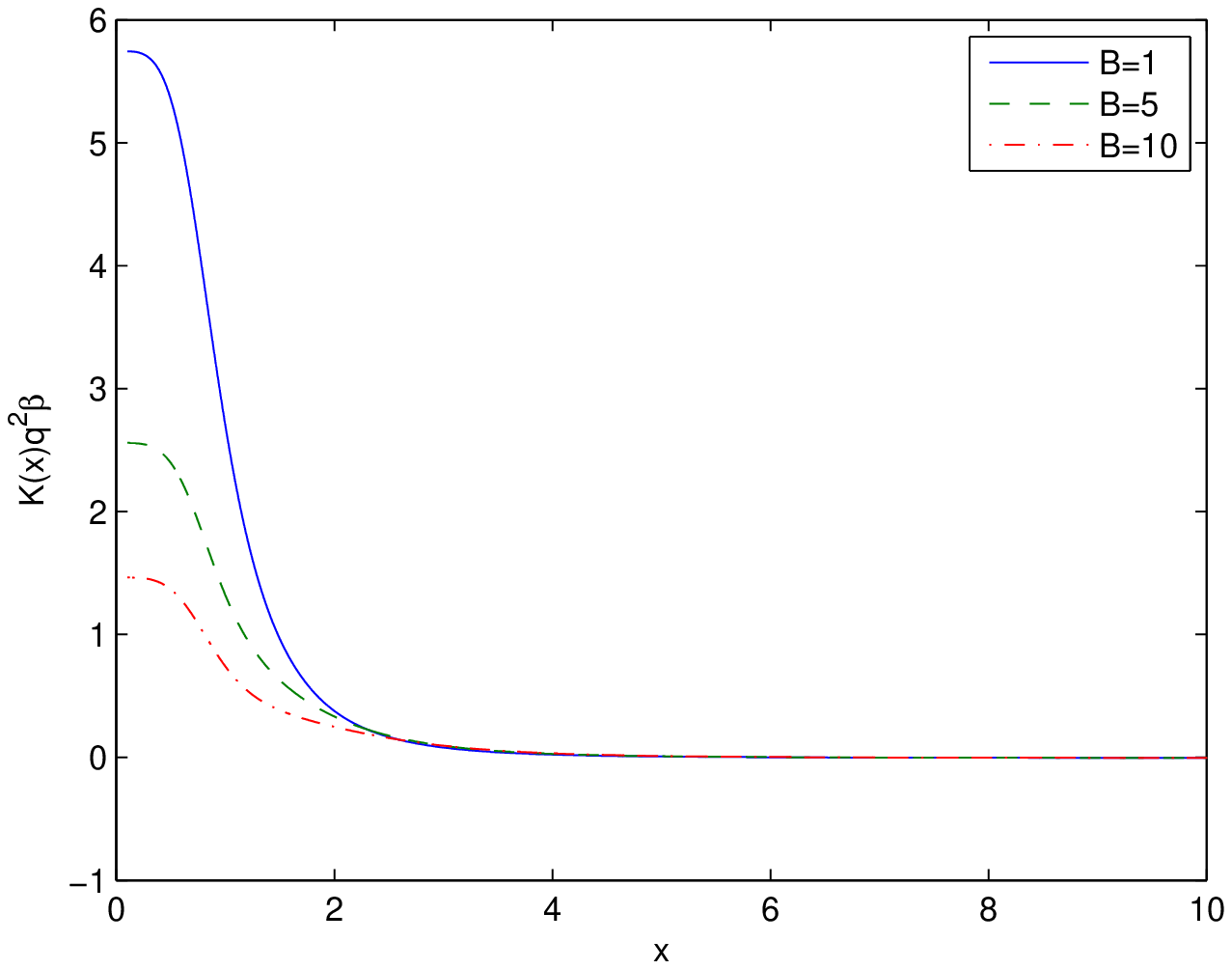}
\caption{\label{fig.6}The plot of the function $K(x)q^2\beta$ versus $x$ for $A=1$, $C=0$ ($m_0=0$). The solid line corresponds to $B=1$, the dashed line corresponds to $B=5$ and the dashed-dotted line corresponds to $B=10$.}
\end{figure}
Figures 5 and 6 show that the curvature invariants are bounded and the limiting curvature conjecture takes place. The maxima of invariants depend on the parameters  $A$, $B$ and $C$. The curvature boundedness is provided by the elementary length $l\neq 0$ in the case of $m_0\neq 0$ for magnetic BH. For the case of $m_0=0$ we have the regular BH in the framework of GR and curvature invariants are bounded.

\section{Conclusion}

The summary of this paper is as follows. We have introduced the fundamental length $l$ which characterises the quantum corrections at short distances close to the Planck length. The classical Einstein theory describes the gravity correctly when the curvature invariants are less than $l^{-2}$. If curvatures are in the order of $l^{-2}$ or greater the quantum effects start to play an important role. In this region the classical GR is UV-incomplete and correct theory should be quantum gravity which is not developed yet. Therefore, in this scale we can describe the quantum effects phenomenologically by modifying the metric function. In this paper we use the Hyward metric for this purpose to take into account quantum corrections to classical Einstein gravity.
For some parameters the magnetically charged BHs have one extreme horizon or two horizons, or no horizons corresponding to the particle-like solution. As the radius approaches infinity corrections to the RN solution  are in the order of ${\cal O}(r^{-4})$ and as $r\rightarrow\infty$ the space-time becomes the Minkowski space-time. As $r\rightarrow 0$ the space-time possesses a de Sitter core and the singularity at $r = 0$ has been smoothed out. Thus, the solution describes the regular BH with the finite curvature everywhere.
By calculating the Hawking temperature and the heat capacity of the BH, we demonstrated that second-order phase transitions occur separating areas between negative and positive heat capacities. When the Hawking temperature is negative the BH does not exist. In some range of horizon radii the heat capacity is positive and the BHs are stable. The behaviors of the Hawking temperature and heat capacity are similar
to them of Ref. [8] and [9]. However, in [8] the expressions for the metric function, temperature and specific heat are given in special functions. In the present paper and in [9] these expressions are simpler. More importantly, the RNED model compared to NED used in [8], [9] has some advantages that we mentioned in Introduction. In addition, it was demonstrated that the limiting curvature conjecture holds.


\begin{thebibliography}{99}

\bibitem{Sakharov} A. D. Sakharov, JETP \textbf{22}, 241 (1966).
\bibitem{Markov} M. A. Markov, JETP Letters \textbf{36}, 266 (1982).
\bibitem{Markov1} M. A. Markov, Ann. Phys. \textbf{155}, 333 (1984).
\bibitem{Frolov} V. P. Frolov, Phys. Rev. D \textbf{94}, 104056 (2016) [arXiv:1609.01758].
\bibitem{Hayward} S. A. Hayward, Phys. Rev. Lett. \textbf{96}, 031103 (2006) [arXiv:gr-qc/0506126].
\bibitem{Bardeen} J. Bardeen, Proc. GR5, Tbilisi, USSR (1968), p.174;
A. Borde, Phys. Rev. D \textbf{50}, 3392 (1994);
C. Barrabes and V. P. Frolov, Phys. Rev. D \textbf{53}, 3215 (1996);
 M. Mars, M. M. Martın–Prats, and J. M. M. Senovilla, Class. Quant. Grav., \textbf{13}, L51 (1996);
 A. Cabo and E. Ayon–Beato, Int. J. Mod. Phys. A \textbf{14}, 2013 (1999);
 A. Borde, Phys. Rev., D \textbf{55}, 7615 (1997).


\bibitem{Kr3} S. I. Kruglov, Ann. Phys. \textbf{353}, 299 (2015) [arXiv:1410.0351].
\bibitem{Ali} A. Ali and K. Saifullah, Phys. Lett. B \textbf{792}, 276 (2019).
\bibitem{Halilsoy} S. H. Mazharimousavi and M. Halilsoy, Phys. Lett. B \textbf{796}, 123 (2019).
\bibitem{Kr8}  S. I. Kruglov, Ann. Phys. \textbf{378}, 59 (2017) [arXiv:1703.02029].
\bibitem{Kr7}  S. I. Kruglov,  Ann. Phys. (Berlin) \textbf{529}, 1700073 (2017) [arXiv:1708.07006].
\bibitem{Kr2}  S. I. Kruglov, Universe \textbf{5}, 225 (2019).
\bibitem{Bronnikov} K. A. Bronnikov, Phys. Rev. D \textbf{63}, 044005 (2001).
\bibitem{BI0} M. Born and L. Infeld, Proc. Roy. Soc. Lond. \textbf{144}, 425 (1934).
\bibitem{BI1} B. Hoffmann, Phys. Rev. \textbf{47}, 877 (1935).
\bibitem{Oliveira} H. P. de Oliveira, Class. Quant. Grav. \textbf{11}, 1469 (1994).
\bibitem{Soleng} H. H. Soleng, Phys. Rev. D \textbf{52}, 6178 (1995) [arXiv:hep-th/9509033].
\bibitem{Kruglov}S.I. Kruglov, Eur. Phys. J. C \textbf{75}, 88 (2015) [arXiv:1411.7741].
\bibitem{Kruglov1}S.I. Kruglov, Ann. Phys. (Berlin) \textbf{527}, 397 (2015) [arXiv:1410.7633].
\bibitem{Ayon-Beato}E. Ayon-Beato and A. Garcia, Phys. Rev. Lett. \textbf{80}, 5056 (1998) [arXiv:gr-qc/9911046].
\bibitem{Bronnikov1} K. A. Bronnikov, Phys. Rev. Lett. \textbf{85}, 4641 (2000).
\bibitem{Hendi} S. H. Hendi, Ann. Phys. \textbf{333}, 282 (2013) [arXiv:1405.5359].
\bibitem{Balart} L. Balart and E. C. Vagenas, Phys. Rev. D \textbf{90}, 124045 (2014) [arXiv:1408.0306].
\bibitem{Kruglov2}S.I. Kruglov, Ann. Phys. \textbf{383}, 550 (2017) [arXiv:1707.04495].
\bibitem{Flachi}A. Flachi and J. P. S. Lemos, Phys. Rev. D \textbf{87}, 024034 (2013) [arXiv:1211.6212].
\bibitem{Habib} I. Gullu, S. H. Mazharimousavi, Phys. Scripta \textbf{96}, 045217 (2021) [arXiv:2009.08665].
 \bibitem{Bronnikov2} K. A. Bronnikov, Int. J. Mod. Phys. D \textbf{27}, 1841005 (2018).
\bibitem{Kr} S. I. Kruglov, Grav. Cosm.  \textbf{27}, 78 (2021) [arXiv:2103.14087].
\bibitem{Kr1} S. I. Kruglov, Mod. Phys. Lett. A \textbf{35}, 2050291 (2020) [arXiv:2009.07657].
\bibitem{Kr4} S. I. Kruglov, Int. J. Mod. Phys. A \textbf{35}, 26 {2020} [arXiv:2009.14637].
\bibitem{Wald}R. M. Wald, General Relativity (University of Chicago Press, 1984).
\bibitem{Hawking} S. W. Hawking, Commun. Math. Phys. \textbf{43}, 199 (1975).
\bibitem{Myung1} Y. S. Myung, Y.-W. Kim, and Y.-J. Park, Phys. Lett. B \textbf{656}, 221 (2007) [arXiv:gr-qc/0702145].
\bibitem{Myung} Y. S. Myung, Y.-W. Kim, and Y.-J. Park, Gen. Rel. Grav. \textbf{41}, 1051 (2009) [arXiv:0708.3145].
\bibitem{Tharanath} R. Tharanath, J. Suresh, and V. C. Kuriakose, Gen. Rel. Grav. \textbf{47}, 46 (2015) [arXiv:1406.3916].
\bibitem{Kr5}  S. I. Kruglov, Universe \textbf{4}, 66 (2018) [arXiv:1805.07595].
\bibitem{Novikov} I. D. Novikov and V. P. Frolov, Physics of Black Holes (Kluver Academic Publishers, 1989).
\bibitem{Polchinski} J. Polchinski,  Nucl. Phys. B  \textbf{325}, 619 (1989).

\end{thebibliography}
\end{document}